\documentclass[a4paper,english,11pt,amssymb,nofootinbib,superscriptaddress,prd]{revtex4}
\usepackage{lmodern}
\usepackage{lmodern}

\usepackage[T1]{fontenc}
\usepackage[latin9]{inputenc}
\usepackage{babel}
\usepackage{amsmath}
\usepackage{amssymb}
\usepackage{esint}
\usepackage[unicode=true,pdfusetitle,
 bookmarks=true,bookmarksnumbered=false,bookmarksopen=false,
 breaklinks=false,pdfborder={0 0 1},backref=false,colorlinks=false]
 {hyperref}

\makeatletter

\@ifundefined{textcolor}{}
{%
 \definecolor{BLACK}{gray}{0}
 \definecolor{WHITE}{gray}{1}
 \definecolor{RED}{rgb}{1,0,0}
 \definecolor{GREEN}{rgb}{0,1,0}
 \definecolor{BLUE}{rgb}{0,0,1}
 \definecolor{CYAN}{cmyk}{1,0,0,0}
 \definecolor{MAGENTA}{cmyk}{0,1,0,0}
 \definecolor{YELLOW}{cmyk}{0,0,1,0}
}

\usepackage{babel}
\usepackage{babel}
\usepackage{babel}
\usepackage{babel}
\usepackage{babel}
\usepackage{babel}
\usepackage{babel}
\usepackage{babel}
\usepackage{babel}
\usepackage{babel}
\usepackage{babel}
\usepackage{babel}
\usepackage{babel}
\usepackage{babel}
\usepackage{babel}
\usepackage{babel}
\usepackage{babel}

\usepackage{babel}
\usepackage{graphicx}
\def\b{\begin{equation}}
\def\e{\end{equation}}

\@ifundefined{textcolor}{}{%
 \definecolor{BLACK}{gray}{0}
 \definecolor{WHITE}{gray}{1}
 \definecolor{RED}{rgb}{1,0,0}
 \definecolor{GREEN}{rgb}{0,1,0}
 \definecolor{BLUE}{rgb}{0,0,1}
 \definecolor{CYAN}{cmyk}{1,0,0,0}
 \definecolor{MAGENTA}{cmyk}{0,1,0,0}
 \definecolor{YELLOW}{cmyk}{0,0,1,0}
 }

\usepackage{latexsym}\usepackage{bm}

\makeatother

\begin{document}
\title{Anomalous Dispersion in Gravity Theories }
\author{{\normalsize{}{}{}{}{}{}{}{}{}{}{}{}{}{}{}{}{}Emel
Altas}}
\email{emelaltas@kmu.edu.tr}

\affiliation{Department of Physics, Karamanoglu Mehmetbey University, 70100, Karaman,
Turkey}
\author{{\normalsize{}{}{}{}{}{}{}{}{}{}{}{}{}{}{}{}{}Ercan
Kilicarslan}}
\email{ercan.kilicarslan@usak.edu.tr}

\affiliation{Department of Mathematics, Usak University, 64200, Usak, Turkey}
\author{{\normalsize{}{}{}{}{}{}{}{}{}{}{}{}{}{}{}{}{}Bayram
Tekin}}
\email{btekin@metu.edu.tr}

\affiliation{Department of Physics, Middle East Technical University, 06800, Ankara,
Turkey}
\date{{\normalsize{}{}{}{}{}{}{}{}{}{}{{}{}{}{}\today}}}

\begin{abstract}
A wave pulse (be it a gravitational wave or a light wave) undergoes anomalous dispersion in a vacuum in flat spacetimes with an even number of spatial dimensions even if all the frequencies move at the same speed.  Such an anomalous dispersion does not occur in spacetimes with an odd number of spatial dimensions. We study various gravity theories and show that dispersion-free propagation is possible in even number of spatial dimensions if the background is not the Minkowski but the de Sitter spacetime and the gravity theory is massive gravity with a tuned mass in terms of the cosmological constant.  Mass and the cosmological constant conspire to get rid of the anomalous dispersion and restore Huygens' principle. 
\end{abstract}

\maketitle
\section{Introduction}

Recently \cite{Tekin_meta} it was shown that the wave equation of a massive Klein-Gordon field with a tuned mass $m= \sqrt{\Lambda}$ in a (2+1)-dimensional de Sitter spacetime, with a positive cosmological $\Lambda$ allows dispersionless propagation. Namely, an initial wave pulse does not broaden and change shape when it propagates in this background. This result, {\it a prirori }, is  counter-intuitive since it is well-known that wave pulses in even spatial dimensions undergo anomalous (dimension-dependent) dispersion even if all modes propagate at the same speed \cite{CH}; and massive wave equations in all dimensions show (regular) dispersion as there is always propagation inside the light-cone due to the fact that the group velocity depends on the wave-number of the individual waves constructing the pulse. But it turns out that these two effects help each other eliminate the anomalous dispersion in certain cases.

The result of \cite{Tekin_meta} was inspired by two works: in \cite{Bender} it was shown that adding one more {\it time-like} dimension to the $(2+1)$ flat spacetime, namely, considering a massless wave equation in a $(2+2)$-dimensional World, one has the possibility of dispersion-free propagation. This is rather surprising since we know that adding a space-like dimension removes the anomalous dispersion, but even an extra time-like direction, albeit physically so removed from a space-like direction, seems to do the job of removing the anomalous dispersion. Even though spacetimes with {\it two time} dimensions appear in theoretical physics \cite{Bars}, one would feel much pleased if an experimental effective model appears to have two time directions. This indeed happens in some hyperbolic metamaterials \cite{SN, Sm2}: in a nondispersive, nonmagnetic, uniaxial anisotropic metamaterial (which can be constructed in a lab) the extraordinary (non-transverse) component of the electric field obeys a {\it massless} Klein-Gordon wave equation in a flat $(2+2)$-dimensional spacetime. In \cite{Bender} this massless wave equation  with constant coefficients in Cartesian coordinates was shown to be  equivalent to a {\it  modified } wave equation with time-dependent coefficients.  Then this modified wave equation was shown to allow particular initial wave pulses 
to propagate without dispersion in vacuum.  The modified wave equation introduced in \cite{Bender} is somewhat {\it ad hoc} and the initial data chosen is rather specific. An explanation was given in \cite{Tekin_meta}: it was shown that the modified wave equation exactly corresponds to a massive scalar field in a $(2+1)$-dimensional de Sitter spacetime  with a tuned mass. Therefore the mentioned hyperbolic metamaterial acts like a $(2+1)$-dimensional de Sitter background and dispersion-free propagation is possible for a generic wave pulse if the mass is tuned to the cosmological constant.  

In this work, building on these considerations, we give a detailed account of propagation of massless and massive gravity waves in generic $D= d+1$ dimensions.  These waves will be gravity waves defined in the weak field limit. It will turn out that in de Sitter backgrounds, massive fields with tuned masses allow dispersion-free propagation generalizing the results of {\cite{Tekin_meta}.

The lay-out of the paper is as follows: In Section II, we study the $D$ dimensional massless gravity (General Relativity) in some detail to set-up the formalism and to see the anomalous dispersion in the behavior of the spacetime Green's functions. In Section III, Fierz-Pauli massive gravity is studied in a flat spacetime background. In Section IV, $D$ dimensional quadratic gravity is studied in a flat spacetime background; and in Section V, 2+1 dimensional topologically massive gravity is studied. In Section VI, 2+1 dimensional new massive gravity and massive Klein-Gordon fields in a $D$-dimensional de Sitter background are studied. The computations are straightforward but rather lengthy, we have provided some of the details of the computations in the appendices.

\section{Massless gravity in $D=d+1$ dimensions }

As it will be our guiding theory, we shall study the $D$ dimensional
massless gravity in some detail. Here we will give the background expansions of the
relevant tensors that will also appear in various massive gravity theories
studied in other sections. In $(d+1)$ dimensions, the Einstein-Hilbert
action reads 
\begin{equation}
I=\frac{1}{2\kappa}\intop d^{d+1}x\thinspace\sqrt{-g}\thinspace R.\label{Einstein-Hilbert action}
\end{equation}
To compute the Green's function of the linearized theory around the
flat spacetime, let us expand the action up to the second order in the
metric fluctuations using 
\begin{equation}
g_{\mu\nu}:=\bar{g}_{\mu\nu}+\tau h_{\mu\nu},\label{metricexpansion}
\end{equation}
where $\tau$ is a small expansion parameter, $\bar{g}_{\mu\nu}$
denotes the flat background spacetime metric in some coordinates.
The inverse metric yields 
\begin{equation}
g^{\mu\nu}=\bar{g}^{\mu\nu}-\tau h^{\mu\nu}+\tau^{2}h^{\mu\sigma}h_{\sigma}^{\nu}+{\mathcal{O}}(\tau^{3}).\label{inversemetricexpansion}
\end{equation}
One also has the expansion of the square root of the determinant of the metric as
\begin{equation}
\sqrt{-g}=\sqrt{-\bar{g}}\left(1+\tau\frac{h}{2}+\tau^{2}\frac{1}{8}(h^{2}-2h_{\mu\nu}^{2})\right),\label{rootgexpansion}
\end{equation}
where $h_{\mu\nu}^{2}=h_{\mu\nu}h^{\mu\nu}$. Expansion of the metric
yields an expansion of tensors that depend on the metric. In particular,
the scalar curvature at the desired order becomes 
\begin{equation}
R=\bar{R}+\tau(R)^{(1)}+\frac{\tau^{2}}{2}(R)^{(2)},\label{scalarcurvatureexpansion}
\end{equation}
where the first and the second order terms can be found to be
\begin{equation}
(R)^{(1)}=\bar{g}^{\mu\nu}(R_{\mu\nu})^{(1)}-h^{\mu\nu}\bar{R}_{\mu\nu},\hskip0.5cm(R)^{(2)}=\bar{g}^{\mu\nu}(R_{\mu\nu})^{(2)}-2h^{\mu\nu}(R_{\mu\nu})^{(1)}+2h^{\mu\sigma}h_{\sigma}^{\nu}\bar{R}_{\mu\nu}.
\end{equation}
The Ricci tensor at the first order can be computed to be 
\begin{equation}
(R_{\mu\nu})^{(1)}=\frac{1}{2}\Big(\bar{\nabla}_{\sigma}\bar{\nabla}_{\mu}h_{\nu}^{\sigma}+\bar{\nabla}_{\sigma}\bar{\nabla}_{\nu}h_{\mu}^{\sigma}-\bar{\square}h_{\mu\nu}-\bar{\nabla}_{\mu}\bar{\nabla}_{\nu}h\Big),\label{linearizedRiccitensor}
\end{equation}
where $\bar{\nabla}_{\mu}$ denotes the background metric compatible
covariant derivative and $\bar{\square}:=\bar{\nabla}_{\mu}\bar{\nabla}^{\mu}$.
So the linearized scalar curvature becomes 
\begin{equation}
(R)^{(1)}=\bar{\nabla}_{\sigma}\bar{\nabla}_{\lambda}h^{\sigma\lambda}-\bar{\square}h-h^{\mu\nu}\bar{R}_{\mu\nu},\label{linearizedscalarcurvature}
\end{equation}
while the second order Ricci tensor is more complicated: 
\begin{equation}
(R_{\mu\nu})^{(2)}=\bar{\nabla}_{\sigma}(\Gamma_{\mu\nu}^{\sigma})^{(2)}-\bar{\nabla}_{\nu}(\Gamma_{\mu\sigma}^{\sigma})^{(2)}+2(\Gamma_{\sigma\lambda}^{\sigma})^{(1)}(\Gamma_{\mu\nu}^{\lambda})^{(1)}-2(\Gamma_{\nu\lambda}^{\sigma})^{(1)}(\Gamma_{\mu\sigma}^{\lambda})^{(1)},
\end{equation}
where $(\Gamma_{\mu\nu}^{\lambda})^{(1)}$ denotes the first order
Christoffel connection that reads as 
\begin{equation}
(\Gamma_{\mu\nu}^{\lambda})^{(1)}=\frac{1}{2}\bar{g}^{\lambda\rho}(\bar{\nabla}_{\mu}h_{\nu\rho}+\bar{\nabla}_{\nu}h_{\mu\rho}-\bar{\nabla}_{\rho}h_{\mu\nu}).
\end{equation}
$(\Gamma_{\mu\nu}^{\sigma})^{(2)}$ is the second order Christoffel
connection of which the explicit form is not needed. Now we can expand
the Einstein-Hilbert action (\ref{Einstein-Hilbert action}) as 
\begin{equation}
I=\bar{I}+\tau(I)^{\left(1\right)}+\frac{\tau^{2}}{2}(I)^{\left(2\right)}.
\end{equation}
After making use of the above results, the second order term boils
down to 
\begin{equation}
(I)^{\left(2\right)}=\frac{1}{2\kappa}\int d^{d+1}x\thinspace\sqrt{-\bar{g}}\thinspace\left(\bar{g}^{\mu\nu}(R_{\mu\nu})^{(2)}-2h^{\mu\nu}(R_{\mu\nu})^{(1)}+2h^{\mu\sigma}h_{\sigma}^{\nu}\bar{R}_{\mu\nu}+h(R)^{(1)}+\frac{1}{4}\bar{R}(h^{2}-2h_{\mu\nu}^{2})\right).\label{secondorderaction}
\end{equation}
This expression is valid for a generic background metric, let us now
consider the flat spacetime with Cartesian coordinates and take $\bar{g}_{\mu\nu}=\eta_{\mu\nu}$,
$\bar{\nabla}_{\mu}=\partial_{\mu}$ and $\bar{R}_{\mu\nu}=0=\bar{R}$.
Then (\ref{secondorderaction}) becomes 
\begin{equation}
(I)^{\left(2\right)}=\frac{1}{2\kappa}\int d^{d+1}x\thinspace\left(\bar{g}^{\mu\nu}(R_{\mu\nu})^{(2)}-2h^{\mu\nu}(R_{\mu\nu})^{(1)}+h(R)^{(1)}\right),
\end{equation}
which, making use of the linearized Einstein tensor 
\begin{equation}
(G_{\mu\nu})^{\left(1\right)}=(R_{\mu\nu})^{\left(1\right)}-\frac{1}{2}\bar{g}_{\mu\nu}(R)^{\left(1\right)}-\frac{1}{2}h_{\mu\nu}\bar{R}=(R_{\mu\nu})^{\left(1\right)}-\frac{1}{2}\bar{g}_{\mu\nu}(R)^{\left(1\right)},\label{linearizedEinsteintensor}
\end{equation}
reduces to 
\begin{equation}
(I)^{\left(2\right)}=\frac{1}{2\kappa}\int d^{d+1}x\thinspace\left(\bar{g}^{\mu\nu}(R_{\mu\nu})^{(2)}-2h^{\mu\nu}(G_{\mu\nu})^{(1)}\right).\label{secondorderactiontwo}
\end{equation}
One can proceed in a gauge-invariant way, but here we impose the harmonic
gauge to simplify the  ensuing expressions. Then assuming 
\begin{equation}
\partial_{\mu}h_{\sigma}^{\mu}=\frac{1}{2}\partial_{\sigma}h,\label{harmonic gauge}
\end{equation}
the linearized Einstein tensor becomes 
\begin{equation}
(G_{\mu\nu})^{(1)}=\frac{1}{4}\Big(\eta_{\mu\nu}\eta_{\alpha\beta}-\eta_{\mu\alpha}\eta_{\nu\beta}-\eta_{\mu\beta}\eta_{\nu\alpha}\Big)\partial^{2}h^{\alpha\beta}.
\end{equation}
Dropping the boundary terms, in the harmonic gauge, one has $\bar{g}^{\mu\nu}(R_{\mu\nu})^{(2)}=h^{\mu\nu}(G_{\mu\nu})^{(1)}$, and 
 (\ref{secondorderactiontwo}) reduces to 
\begin{equation}
(I)^{\left(2\right)}=-\frac{1}{2\kappa}\int d^{d+1}x\thinspace h^{\mu\nu}(G_{\mu\nu})^{(1)}.\label{secondorderactionthree}
\end{equation}
This can be written as 
\begin{equation}
(I)^{\left(2\right)}=\frac{1}{4\kappa}\int d^{d+1}xh^{\mu\nu}{\cal O}_{\mu\nu\alpha\beta}(x)h^{\alpha\beta},\label{Etics1}
\end{equation}
with the formally self-adjoint operator given as 
\begin{equation}
{\cal O}_{\mu\nu\alpha\beta}(x)=-\frac{1}{2}\Big(\eta_{\mu\nu}\eta_{\alpha\beta}-\eta_{\mu\alpha}\eta_{\nu\beta}-\eta_{\mu\beta}\eta_{\nu\alpha}\Big)\partial^{2}.\label{operator}
\end{equation}
Green's function is the inverse of the operator ${\cal O}_{\mu\nu\alpha\beta}$,
under the assumed (sufficient decay at infinity)  boundary conditions, hence one must solve the equation
\begin{equation}
{\cal O}_{\mu\nu\alpha\beta}(x)G^{\alpha\beta\lambda\tau}(x,x')=\frac{1}{2}(\delta_{\mu}^{\lambda}\delta_{\nu}^{\tau}+\delta_{\nu}^{\lambda}\delta_{\mu}^{\tau})\delta^{(d+1)}(x-x'),
\end{equation}
which, in the momentum space, reads as 
\begin{equation}
\tilde{{\cal {O}}}_{\mu\nu\alpha\beta}(p)\,\tilde{G}^{\alpha\beta\lambda\tau}(p)=\frac{1}{2}(\delta_{\mu}^{\lambda}\delta_{\nu}^{\tau}+\delta_{\nu}^{\lambda}\delta_{\mu}^{\tau}).\label{identity}
\end{equation}
${\cal \tilde{O}}_{\mu\nu\alpha\beta}$ can be obtained from (\ref{operator})
by replacing $\partial_{\mu}$ with $ip_{\mu}$ to get 
\begin{equation}
{\cal \tilde{O}}_{\mu\nu\alpha\beta}(p)=\frac{p^{2}}{2}\Big(\eta_{\mu\nu}\eta_{\alpha\beta}-\eta_{\mu\alpha}\eta_{\nu\beta}-\eta_{\mu\beta}\eta_{\nu\alpha}\Big).
\end{equation}
Then the solution satisfying (\ref{identity}) is 
\begin{equation}
\tilde{G}^{\alpha\beta\lambda\tau}(p)=-\frac{1}{2p^{2}}\Big(\eta^{\alpha\lambda}\eta^{\beta\tau}+\eta^{\alpha\tau}\eta^{\beta\lambda}-\frac{2\eta^{\alpha\beta}\eta^{\lambda\tau}}{d-1}\Big).\label{propagator}
\end{equation}
The position space Green's function can be obtained from the Fourier
transform 
\begin{equation}
G^{\alpha\beta\lambda\tau}(x,x')=\int\frac{d^{d+1}p}{(2\pi)^{d+1}}e^{-ip\cdot(x-x')}\tilde{G}^{\alpha\beta\lambda\tau}(p),
\end{equation}
which reads 
\begin{equation}
G^{\alpha\beta\lambda\tau}(x,x')=-\frac{1}{2}\Big(\eta^{\alpha\lambda}\eta^{\beta\tau}+\eta^{\alpha\tau}\eta^{\beta\lambda}-\frac{2\eta^{\alpha\beta}\eta^{\lambda\tau}}{d-1}\Big)\int\frac{d^{d+1}p}{(2\pi)^{d+1}}e^{-ip\cdot(x-x')}\frac{1}{p^{2}}.
\end{equation}
We are looking for the retarded Green's function, the poles should
be displaced as such, the result of the integral depends on the number
of dimensions. For $d\ge2$, defining $t:=t-t'$ and $r:=|\vec{x}-\vec{x}'|$,
one arrives at (see the Appendix for details) 
\begin{eqnarray}
G^{\alpha\beta\lambda\tau}(t,r)=-\frac{1}{2}\Big(\eta^{\alpha\lambda}\eta^{\beta\tau}+\eta^{\alpha\tau}\eta^{\beta\lambda}-\frac{2\eta^{\alpha\beta}\eta^{\lambda\tau}}{d-1}\Big)\left\{ \begin{array}{lr}
\frac{1}{4\pi}\Big(-\frac{1}{2\pi r}\partial_{r}\Big)^{\frac{d-3}{2}}\frac{\delta(t-r)}{r} & :\,\text{for}\,\,\text{odd}\,\,d,\\
\\
\frac{\theta(t)}{2\pi}\Big(-\frac{1}{2\pi r}\partial_{r}\Big)^{\frac{d}{2}-1}\frac{\theta(t-r)}{\sqrt{t^{2}-r^{2}}} & :\,\text{for}\,\text{even}\,\,d.
\end{array}\right.\label{Greenilk}
\end{eqnarray}
For odd $d$, the Green's function is non-zero only for null separation
and hence there is no tail inside the light-cone. On the other hand,
for even $d$, even though the Green's function is peaked around for
the null separation due to the appearance of the function $\frac{1}{\sqrt{t^{2}-r^{2}}}$,
there is a tail inside the light cone. Hence, even a delta-function initial wave is dispersed and one has anomalous dispersion of gravitational
waves. Note that, among the odd spatial dimensions, $d=3$, our World is special since
only for this dimension, there is no derivative on the delta function,
hence the delta function pulse at $t=0$ remains a delta function, only-shifted to a new location, at 
all points and for all times.

\section{Fierz-Pauli Massive Gravity in $D=d+1$ dimensions}

The linearized Fierz-Pauli action, that has $\frac{(D+1)(D-2)}{2}$
degrees of freedom in a flat spacetime background, is 
\begin{equation}
I=\frac{1}{2\kappa}\int d^{d+1}x\left(-\frac{1}{2}\partial_{\lambda}h^{\mu\nu}\partial^{\lambda}h_{\mu\nu}+\partial^{\nu}h^{\lambda\mu}\partial_{\mu}h_{\lambda\nu}-\partial_{\mu}h^{\mu\nu}\partial_{\nu}h+\frac{1}{2}\partial_{\lambda}h\partial^{\lambda}h-\frac{m^{2}}{2}(h_{\mu\nu}^{2}-h^{2})\right).\label{FP}
\end{equation}
The field equations coming from this action 
\begin{equation}
\partial^{2}h_{\mu\nu}-\partial_{\lambda}\partial_{\nu}h_{\mu}^{\lambda}-\partial_{\lambda}\partial_{\mu}h_{\nu}^{\lambda}+\partial_{\nu}\partial_{\mu}h+\eta_{\mu\nu}\partial_{\sigma}\partial_{\lambda}h^{\sigma\lambda}-\eta_{\mu\nu}\partial^{2}h=m^{2}(h_{\mu\nu}-\eta_{\mu\nu}h)
\end{equation}
can be recast as three equations 
\begin{equation}
(\partial^{2}-m^{2})h_{\mu\nu}=0,~~~~~~~~\partial_{\mu}h^{\mu\nu}=0,~~~~~~~~h=0.
\end{equation}
The action (\ref{FP}) up to boundary terms reads 
\begin{equation}
I=\frac{1}{4\kappa}\int d^{d+1}xh^{\mu\nu}{\cal O}_{\mu\nu\alpha\beta}(x)h^{\alpha\beta},\label{Etics1}
\end{equation}
with 
\begin{eqnarray}
 &  & {\cal O}_{\mu\nu\alpha\beta}(x)=\frac{1}{2}\bigl(\eta_{\mu\alpha}\eta_{\nu\beta}+\eta_{\mu\beta}\eta_{\nu\alpha}\bigr)(\partial^{2}-m^{2})+\eta_{\alpha\beta}\bigl(\partial_{\mu}\partial_{\nu}-\eta_{\mu\nu}(\partial^{2}-m^{2})\bigr)+\eta_{\mu\nu}\partial_{\alpha}\partial_{\beta}\nonumber \\
 &  & \ ~\ ~\ ~\ ~\ ~\ ~-\frac{1}{2}\bigl(\eta_{\mu\beta}\partial_{\alpha}\partial_{\nu}+\eta_{\mu\alpha}\partial_{\alpha}\partial_{\nu}+\eta_{\nu\beta}\partial_{\alpha}\partial_{\mu}+\eta_{\nu\alpha}\partial_{\alpha}\partial_{\mu}\bigr).
\end{eqnarray}
Following the steps in the previous section verbatim, one arrives
at the momentum space Green's function
\begin{equation}
\tilde{G}^{\alpha\beta\sigma\lambda}(p)=-\frac{1}{2(p^{2}+m^{2})}\Big(\eta^{\alpha\sigma}\eta^{\beta\lambda}+\eta^{\alpha\lambda}\eta^{\beta\sigma}-\frac{2}{d}\eta^{\alpha\beta}\eta^{\sigma\lambda}\Big),
\end{equation}
in which we dropped the terms proportional to $p^{\alpha}$ \textit{etc.}
as they do not contribute to any calculation for which the energy-momentum
tensor is conserved ($p^\alpha T_{ \alpha \beta}=0$). Then we have the position space Green's function
\begin{equation}
G^{\alpha\beta\sigma\lambda}(x,x')=-\frac{1}{2}\Big(\eta^{\alpha\sigma}\eta^{\beta\lambda}+\eta^{\alpha\lambda}\eta^{\beta\sigma}-\frac{2}{d}\eta^{\alpha\beta}\eta^{\sigma\lambda}\Big)\int\frac{d^{d+1}p}{(2\pi)^{d+1}}\frac{e^{-ip(x-x')}}{p^{2}+m^{2}}.
\end{equation}
Once again the results of this integral differ for odd and even $d$
(see the Appendix for discussion). For odd $d$, one has 
\begin{equation}
G_{{\mbox{odd}}\,d}^{\alpha\beta\sigma\lambda}(t,r)=-\frac{1}{2}\Big(\eta^{\alpha\sigma}\eta^{\beta\lambda}+\eta^{\alpha\lambda}\eta^{\beta\sigma}-\frac{2}{d}\eta^{\alpha\beta}\eta^{\sigma\lambda}\Big)\frac{\Theta(t)}{2}\Bigg(-\frac{1}{2\pi r}\frac{d}{dr}\Bigg)^{\frac{d-1}{2}}\Bigg(J_{0}(m\sqrt{t^{2}-r^{2}})\Theta(t-r)\Bigg),
\end{equation}
where $J_0$ is the Bessel function. For example for $d=3$, one gets 
\begin{equation}
G^{\alpha\beta\sigma\lambda}(t,r)=-\frac{1}{2}\Bigg(\eta^{\alpha\sigma}\eta^{\beta\lambda}+\eta^{\alpha\lambda}\eta^{\beta\sigma}-\frac{2}{3}\eta^{\alpha\beta}\eta^{\sigma\lambda}\Bigg)\frac{\Theta(t)}{2}\Bigg(-\frac{1}{2\pi r}\frac{d}{dr}\Bigg)\Bigg(J_{0}(m\sqrt{t^{2}-r^{2}})\Theta(t-r)\Bigg).\label{d3case}
\end{equation}
The derivative part yields 
\begin{equation}
\Bigg(-\frac{1}{2\pi r}\frac{d}{dr}\Bigg)\Bigg(J_{0}(m\sqrt{t^{2}-r^{2}})\Theta(t-r)\Bigg)=\frac{\delta(t-r)J_{0}\left(m\sqrt{t^{2}-r^{2}}\right)}{2\pi r}-\frac{m\theta(t-r)J_{1}\left(m\sqrt{t^{2}-r^{2}}\right)}{2\pi\sqrt{t^{2}-r^{2}}}.
\end{equation}
In the $m\rightarrow0$ limit, this last equation gives the expected result  $\frac{\delta(t-r)}{2\pi r}$; but (\ref{d3case}) does not smoothly reduce to the corresponding
$d=3$ case of (\ref{Greenilk}) due to the discrete difference in
the third terms in the first brackets. This is the well-known van-Dam-Veltman-Zakharov
discontinuity. Generically, as expected, in flat space for non-zero
$m$, there is a tail inside the light-cone and the retarded Green's
function has support inside the light-cone.

For even $d$, one has 
\begin{eqnarray}
G_{{\mbox{even}}\,d}^{\alpha\beta\sigma\lambda}(t,r)=-\frac{1}{2}(\eta^{\alpha\sigma}\eta^{\beta\lambda}+\eta^{\alpha\lambda}\eta^{\beta\sigma}-\frac{2}{d}\eta^{\alpha\beta}\eta^{\sigma\lambda})\frac{\Theta(t)}{2\pi}\Big(\frac{-1}{2\pi r}\frac{d}{dr}\Big)^{\frac{d-2}{2}}(\frac{\cos(m\sqrt{t^{2}-r^{2}})\Theta(t-r)}{\sqrt{t^{2}-r^{2}}}).
\end{eqnarray}
For $d=2$, this yields 
\begin{equation}
G^{\alpha\beta\sigma\lambda}(t,r)=-\frac{1}{2}\Bigg(\eta^{\alpha\sigma}\eta^{\beta\lambda}+\eta^{\alpha\lambda}\eta^{\beta\sigma}-\eta^{\alpha\beta}\eta^{\sigma\lambda}\Bigg)\frac{\Theta(t)}{2\pi}\cos(m\sqrt{t^{2}-r^{2}})\frac{\Theta(t-r)}{\sqrt{t^{2}-r^{2}}}.
\end{equation}
For the even $d$ case, there is a support inside the light-cone and
the Huygens' principle is violated. These results are expected in
flat spacetime for massive fields.

\section{Quadratic Curvature Gravity in $D=d+1$ dimensions}

We consider the following quadratic gravity action~\footnote{We do not consider the $R_{\mu\nu\alpha\beta}^{2}$ term, since at
the end we would like to study the particular 2+1 dimensional gravity
for which this term only shifts the parameters in the Lagrangian.} 
\begin{equation}
I_{quad}=\frac{1}{2\kappa}\int d^{d+1}x\sqrt{-g}\bigg(\sigma R+\alpha R^{2}+\beta R_{\mu\nu}^{2}\bigg),\label{QGaction}
\end{equation}
from which the second order action in the harmonic gauge can be found to be 
\begin{equation}
(I_{quad})^{\left(2\right)}=\frac{1}{4\kappa}\int d^{d+1}x\thinspace h^{\mu\nu}{\cal O}_{\mu\nu\alpha\beta}(x)h^{\alpha\beta},
\end{equation}
where the inverse propagator is a fourth order operator 
\begin{equation}
{\cal O}_{\mu\nu\alpha\beta}(x)=\eta_{\mu\nu}\eta_{\alpha\beta}\Big((2\alpha+\frac{\beta}{2})\partial^{2}-\frac{\sigma}{2}\Big)\partial^{2}-(2\alpha+\beta)\eta_{\alpha\beta}\partial_{\mu}\partial_{\nu}\partial^{2}+\frac{1}{2}\Big(\eta_{\mu\alpha}\eta_{\nu\beta}+\eta_{\mu\beta}\eta_{\nu\alpha}\Big)(\sigma+\beta\partial^{2})\partial^{2}.
\end{equation}
From the Fourier transform of this operator, one can find the Green's
function in the momentum space following the similar steps as in 
the second section. The Green's function satisfying (\ref{identity})
reads in the momentum space as 
\begin{eqnarray}
 &  & \tilde{G}^{\alpha\beta\lambda\tau}(p)=\Bigl(-\frac{1}{2\sigma}(\eta^{\alpha\lambda}\eta^{\beta\tau}+\eta^{\alpha\tau}\eta^{\beta\lambda})+\frac{1}{\sigma(d-1)}\eta^{\alpha\beta}\eta^{\lambda\tau}-\frac{4\alpha+2\beta}{\sigma^{2}(d-1)}\eta^{\lambda\tau}p^{\beta}p^{\alpha}\Bigr)\frac{1}{p^{2}}\nonumber \\
 &  & ~~~~~~~~~~~~~~~~+\Bigl(\frac{1}{2\sigma}(\eta^{\alpha\lambda}\eta^{\beta\tau}+\eta^{\alpha\tau}\eta^{\beta\lambda})-\frac{1}{\sigma d}\eta^{\alpha\beta}\eta^{\lambda\tau}+\frac{\beta}{\sigma^{2}d}\eta^{\lambda\tau}p^{\beta}p^{\alpha}\Bigr)\frac{1}{p^{2}-\sigma/\beta}\label{propagatorquadraticgravity}\\
 &  & ~~~~~~~~~~~~~~~~+\Bigl(-\frac{1}{\sigma d(d-1)}\eta^{\alpha\beta}\eta^{\lambda\tau}+\frac{4\alpha d+\beta(d+1)}{\sigma^{2}d(d-1)}\eta^{\lambda\tau}p^{\beta}p^{\alpha}\Bigr)\frac{1}{p^{2}+\frac{\sigma(d-1)}{4\alpha d+\beta(d+1)}}.\nonumber 
\end{eqnarray}
From the pole structure, one can read the masses of the excitations:
there is a massless spin 2 particle, there is a massive spin 2 particle with
a mass $m_{g}^{2}=-\frac{\sigma}{\beta}$, 
and there is a massive scalar mode with a mass $m_{s}^{2}=\frac{\sigma(d-1)}{4\alpha d+\beta(d+1)}$. A complimentary study of these in (anti) de Sitter spacetimes can be found in \cite{Tekin_Rapid}:
the masses get non-trivial contributions from the non-zero constant
curvature background. In generic $D$-dimensions, this theory has a massive ghost \cite{Stelle} which only disappears for $D=3$ in a particular tuning of $\alpha$ and $\beta$ which we shall study in the next section. 

To get the retarded Green's function in the position space, we have
to do the following integrals 
\begin{eqnarray}
 &  & G^{\alpha\beta\lambda\tau}(x,x')=\Bigl(-\frac{1}{2\sigma}(\eta^{\alpha\lambda}\eta^{\beta\tau}+\eta^{\alpha\tau}\eta^{\beta\lambda})+\frac{1}{\sigma(d-1)}\eta^{\alpha\beta}\eta^{\lambda\tau}+\frac{4\alpha+2\beta}{\sigma^{2}(d-1)}\eta^{\lambda\tau}\partial^{\beta}\partial^{\alpha}\Bigr)\nonumber \\
&& \hskip 5 cm \times \int\frac{d^{d+1}p}{(2\pi)^{d+1}}\frac{e^{-ip(x-x')}}{p^{2}} \\
 &  & ~~~~~~~~~~~~+\Bigl(\frac{1}{2\sigma}(\eta^{\alpha\lambda}\eta^{\beta\tau}+\eta^{\alpha\tau}\eta^{\beta\lambda})-\frac{1}{\sigma d}\eta^{\alpha\beta}\eta^{\lambda\tau}-\frac{\beta}{\sigma^{2}d}\eta^{\lambda\tau}\partial^{\beta}\partial^{\alpha}\Bigr)\int\frac{d^{d+1}p}{(2\pi)^{d+1}}\frac{e^{-ip(x-x')}}{p^{2}-\sigma/\beta} \nonumber\\
 &  & ~~~~~~~~~~~~~-\Bigl(\frac{1}{\sigma d(d-1)}\eta^{\alpha\beta}\eta^{\lambda\tau}+\frac{4\alpha d+\beta(d+1)}{\sigma^{2}d(d-1)}\eta^{\lambda\tau}\partial^{\beta}\partial^{\alpha}\Bigr)\int\frac{d^{d+1}p}{(2\pi)^{d+1}}\frac{e^{-ip(x-x')}}{p^{2}+\frac{\sigma(d-1)}{4\alpha d+\beta(d+1)}},\nonumber 
\end{eqnarray}
which again should be studied in odd and even $d$ separately.

\textbf{i: Odd $d$ case}

\begin{eqnarray}
 &  & G^{\alpha\beta\lambda\tau}(t,r)=\Bigl(-\frac{1}{2\sigma}(\eta^{\alpha\lambda}\eta^{\beta\tau}+\eta^{\alpha\tau}\eta^{\beta\lambda})+\frac{1}{\sigma(d-1)}\eta^{\alpha\beta}\eta^{\lambda\tau}+\frac{4\alpha+2\beta}{\sigma^{2}(d-1)}\eta^{\lambda\tau}\partial^{\beta}\partial^{\alpha}\Bigr)\\
 &  & ~~~~~~~~~~~~~~~~~\times\frac{1}{4\pi}\Theta(t)\bigl(-\frac{1}{2\pi r}\frac{d}{dr}\bigr)^{(d-3)/2}\frac{\delta(t-r)}{r}\nonumber \\
 &  & ~~~~~~~~~~~~+\Bigl(\frac{1}{2\sigma}(\eta^{\alpha\lambda}\eta^{\beta\tau}+\eta^{\alpha\tau}\eta^{\beta\lambda})-\frac{1}{\sigma d}\eta^{\alpha\beta}\eta^{\lambda\tau}-\frac{\beta}{\sigma^{2}d}\eta^{\lambda\tau}\partial^{\beta}\partial^{\alpha}\Bigr)\nonumber \\
 &  & ~~~~~~~~~~~~~~~~~\times\frac{1}{2}\Theta(t)\bigl(-\frac{1}{2\pi r}\frac{d}{dr}\bigr)^{(d-1)/2}J_{0}\Bigl(m_{g}\sqrt{t^{2}-r^{2}}\Bigr)\Theta(t-r)\nonumber \\
 &  & ~~~~~~~~~~~~~-\Bigl(\frac{1}{\sigma d(d-1)}\eta^{\alpha\beta}\eta^{\lambda\tau}+\frac{1}{\sigma dm_{s}^{2}}\eta^{\lambda\tau}\partial^{\beta}\partial^{\alpha}\Bigr)\nonumber \\
 &  & ~~~~~~~~~\times\frac{1}{2}\Theta(t)\Bigl(-\frac{1}{2\pi r}\frac{d}{dr}\Bigr)^{(d-1)/2}J_{0}\Bigl(m_{s}\sqrt{t^{2}-r^{2}}\Bigr)\Theta(t-r),\nonumber 
\end{eqnarray}
where we have used the explicit forms of the masses $m_{g}$ and $m_{s}$.
In particular for $d=3$, one arrives at 
\begin{eqnarray}
G^{\alpha\beta\lambda\tau}(t,r)= &  & \Bigl(-\frac{1}{2\sigma}(\eta^{\alpha\lambda}\eta^{\beta\tau}+\eta^{\alpha\tau}\eta^{\beta\lambda})+\frac{1}{2\sigma}\eta^{\alpha\beta}\eta^{\lambda\tau}+\frac{4\alpha+2\beta}{2\sigma^{2}}\eta^{\lambda\tau}\partial^{\beta}\partial^{\alpha}\Bigr)\frac{1}{4\pi}\frac{\Theta(t)\delta(t-r)}{r}\nonumber \\
 &  & +\Bigl(\frac{1}{2\sigma}(\eta^{\alpha\lambda}\eta^{\beta\tau}+\eta^{\alpha\tau}\eta^{\beta\lambda})-\frac{1}{3\sigma}\eta^{\alpha\beta}\eta^{\lambda\tau}-\frac{\beta}{3\sigma^{2}}\eta^{\lambda\tau}\partial^{\beta}\partial^{\alpha}\Bigr)\\
 &  & ~~~~~~~~~~~~~~\times\frac{\Theta(t)}{2}\Bigl(-\frac{1}{2\pi r}\frac{d}{dr}\Bigr)J_{0}\Bigl(m_{g}\sqrt{t^{2}-r^{2}}\Bigr)\Theta(t-r)\nonumber \\
 &  & ~-\Bigl(\frac{1}{6\sigma}\eta^{\alpha\beta}\eta^{\lambda\tau}+\frac{12\alpha+4\beta}{6\sigma^{2}}\eta^{\lambda\tau}\partial^{\beta}\partial^{\alpha}\Bigr)\times\frac{\Theta(t)}{2}\Bigl(-\frac{1}{2\pi r}\frac{d}{dr}\Bigr)J_{0}\Bigl(m_{s}\sqrt{t^{2}-r^{2}}\Bigr)\Theta(t-r).\nonumber 
\end{eqnarray}
Again, due to the massive parts, as expected, there is propagation
inside the light-cone. Hence the quadratic gravity violates the Huygens'
principle in a flat spacetime.
\newpage 
\textbf{ii: Even $d$ case}

\begin{eqnarray}
 &  & G^{\alpha\beta\lambda\tau}(t,r)=\Bigl(-\frac{1}{2\sigma}(\eta^{\alpha\lambda}\eta^{\beta\tau}+\eta^{\alpha\tau}\eta^{\beta\lambda})+\frac{1}{\sigma(d-1)}\eta^{\alpha\beta}\eta^{\lambda\tau}+\frac{4\alpha+2\beta}{\sigma^{2}(d-1)}\eta^{\lambda\tau}\partial^{\beta}\partial^{\alpha}\Bigr)\\
 &  & ~~~~~~~~~~~~~~~~~\times\frac{1}{2\pi}\Theta(t)\Bigl(-\frac{1}{2\pi r}\frac{d}{dr}\Bigr)^{(d-2)/2}\frac{\Theta(t-r)}{\sqrt{t^{2}-r^{2}}}\nonumber \\
 &  & ~~~~~~~~~~~~+\Bigl(\frac{1}{2\sigma}(\eta^{\alpha\lambda}\eta^{\beta\tau}+\eta^{\alpha\tau}\eta^{\beta\lambda})-\frac{1}{\sigma d}\eta^{\alpha\beta}\eta^{\lambda\tau}-\frac{\beta}{\sigma^{2}d}\eta^{\lambda\tau}\partial^{\beta}\partial^{\alpha}\Bigr)\nonumber \\
 &  & ~~~~~~~~~~~~~~~~~\times\frac{1}{2\pi}\Theta(t)\Bigl(-\frac{1}{2\pi r}\frac{d}{dr}\Bigr)^{(d-2)/2}\cos\Bigl(m_{g}\sqrt{t^{2}-r^{2}}\Bigr)\frac{\Theta(t-r)}{\sqrt{t^{2}-r^{2}}}\nonumber \\
 &  & ~~~~~~~~~~~~~-\Bigl(\frac{1}{\sigma d(d-1)}\eta^{\alpha\beta}\eta^{\lambda\tau}+\frac{1}{\sigma dm_{s}^{2}}\eta^{\lambda\tau}\partial^{\beta}\partial^{\alpha}\Bigr)\nonumber \\
 &  & ~~~~~~~~~\times\frac{1}{2\pi}\Theta(t)\Bigl(-\frac{1}{2\pi r}\frac{d}{dr}\Bigr)^{(d-2)/2}\cos\Bigl(m_{s}\sqrt{t^{2}-r^{2}}\Bigr)\frac{\Theta(t-r)}{\sqrt{t^{2}-r^{2}}}.\nonumber 
\end{eqnarray}
In particular, for $d=2$, one has 
\begin{eqnarray}
 &  & G^{\alpha\beta\lambda\tau}(t,r)=\Bigl(-\frac{1}{2\sigma}(\eta^{\alpha\lambda}\eta^{\beta\tau}+\eta^{\alpha\tau}\eta^{\beta\lambda})+\frac{1}{\sigma}\eta^{\alpha\beta}\eta^{\lambda\tau}+\frac{4\alpha+2\beta}{\sigma^{2}}\eta^{\lambda\tau}\partial^{\beta}\partial^{\alpha}\Bigr)\frac{1}{2\pi}\Theta(t)\frac{\Theta(t-r)}{\sqrt{t^{2}-r^{2}}}\nonumber \\
 &  & +\Bigl(\frac{1}{2\sigma}(\eta^{\alpha\lambda}\eta^{\beta\tau}+\eta^{\alpha\tau}\eta^{\beta\lambda})-\frac{1}{2\sigma}\eta^{\alpha\beta}\eta^{\lambda\tau}-\frac{\beta}{2\sigma^{2}}\eta^{\lambda\tau}\partial^{\beta}\partial^{\alpha}\Bigr)\frac{\Theta(t)}{2\pi}\cos\Bigl(i\sqrt{\sigma/\beta}\sqrt{t^{2}-r^{2}}\Bigr)\frac{\Theta(t-r)}{\sqrt{t^{2}-r^{2}}}\nonumber \\
 &  & -\Bigl(\frac{1}{2\sigma}\eta^{\alpha\beta}\eta^{\lambda\tau}+\frac{8\alpha+3\beta}{2\sigma^{2}}\eta^{\lambda\tau}\partial^{\beta}\partial^{\alpha}\Bigr)\frac{\Theta(t)}{2\pi}\cos\Bigl(\sqrt{\frac{\sigma}{8\alpha+3\beta}}\sqrt{t^{2}-r^{2}}\Bigr)\frac{\Theta(t-r)}{\sqrt{t^{2}-r^{2}}}.\label{genelquad2d}
\end{eqnarray}

A particular $2+1$ dimensional model, the so called New Massive Gravity
(NMG) \cite{nmg1,nmg2,Gullu} is one of our main interests here. So let
us consider this theory. Choosing $\beta=1/m^{2}$
and $\alpha=-3/(8m^{2})$ and $\sigma=-1$, ({\ref{genelquad2d})
yields 
\begin{eqnarray}
 &  & G_{\text{NMG}}^{\alpha\beta\lambda\tau}(t,r)=\frac{1}{4\pi}\left(\eta^{\alpha\lambda}\eta^{\beta\tau}+\eta^{\alpha\tau}\eta^{\beta\lambda}-2\eta^{\alpha\beta}\eta^{\lambda\tau}+\frac{1}{m^{2}}\eta^{\lambda\tau}\partial^{\beta}\partial^{\alpha}\right)\frac{\Theta(t)\Theta(t-r)}{\sqrt{t^{2}-r^{2}}}\\
 &  & -\frac{1}{4\pi}\left(\eta^{\alpha\lambda}\eta^{\beta\tau}+\eta^{\alpha\tau}\eta^{\beta\lambda}-\eta^{\alpha\beta}\eta^{\lambda\tau}+\frac{1}{m^{2}}\eta^{\lambda\tau}\partial^{\beta}\partial^{\alpha}\right)\frac{\cos(m\sqrt{t^{2}-r^{2}})}{\sqrt{t^{2}-r^{2}}}\varTheta(t)\varTheta(t-r).\nonumber 
\end{eqnarray}
There is propagation inside the light-cone and hence NMG in flat spacetime
violates the Huygens' principle. We shall come back to the de Sitter version of this theory in Section VI.

\section{Topologically Massive Gravity}

The action for TMG is \cite{djt}
\begin{equation}
I_{TMG}=\intop d^{3}x\thinspace\sqrt{-g}\left(\frac{1}{\kappa}R+\frac{1}{2\mu}\,\eta^{\mu\nu\alpha}\Gamma^{\beta}{_{\mu\sigma}}\Big(\partial_{\nu}\Gamma^{\sigma}{_{\alpha\beta}}+\frac{2}{3}\Gamma^{\sigma}{_{\nu\lambda}}\Gamma^{\lambda}{_{\alpha\beta}}\Big)\right),\label{wtics1}
\end{equation}
where $\eta^{\mu\nu\alpha}$ is the $3D$ antisymmetric tensor. The action yields
the following field equations 
\begin{equation}
\frac{1}{\kappa}G_{\mu\nu}+\frac{1}{\mu}C_{\mu\nu}=0,\label{tmgfeq}
\end{equation}
where $C_{\mu\nu}$ denotes the Cotton tensor given as 
\begin{equation}
C_{\mu\nu}=\eta_{\mu}\thinspace^{\sigma\rho}\nabla_{\sigma}\bigl(R_{\rho\nu}-\frac{1}{4}g_{\rho\nu}R\bigr),\label{cotton3}
\end{equation}
which is symmetric, divergence-free and traceless. Linearization of
the action around the flat spacetime yields the inverse propagator
\begin{eqnarray}
 &  & {\cal O}_{\mu\nu\alpha\beta}(x)=-\frac{1}{2\kappa}\Bigl(\eta_{\mu\nu}\eta_{\alpha\beta}-\eta_{\mu\alpha}\eta_{\nu\beta}-\eta_{\mu\beta}\eta_{\nu\alpha}\Bigr)\partial^{2}\\
 &  & ~~~~~~~~~~~~+\frac{1}{4\mu}\Bigl(\eta_{\mu}\thinspace^{\sigma}\thinspace_{\alpha}\eta_{\nu\beta}+\eta_{\mu}\thinspace^{\sigma}\thinspace_{\beta}\eta_{\nu\alpha}+\eta_{\nu}\thinspace^{\sigma}\thinspace_{\alpha}\eta_{\mu\beta}+\eta_{\nu}\thinspace^{\sigma}\thinspace_{\beta}\eta_{\mu\alpha}\Bigr)\partial_{\sigma}\partial^{2},\nonumber 
\end{eqnarray}
which in momentum space becomes
\begin{eqnarray}
 &  & {\cal \widetilde{O}}_{\mu\nu\alpha\beta}(p)=\frac{p^{2}}{2\kappa}\Bigl(\eta_{\mu\nu}\eta_{\alpha\beta}-\eta_{\mu\alpha}\eta_{\nu\beta}-\eta_{\mu\beta}\eta_{\nu\alpha}\Bigr)\\
 &  & ~~~~~~~~~~~~-\frac{p^{2}}{4\mu}p^{\lambda}\Bigl(\eta_{\mu\lambda\alpha}\eta_{\nu\beta}+\eta_{\mu\lambda\beta}\eta_{\nu\alpha}+\eta_{\nu\lambda\alpha}\eta_{\mu\beta}+\eta_{\nu\lambda\beta}\eta_{\mu\alpha}\Bigr).\nonumber 
\end{eqnarray}
We obtain  the momentum space propagator as
\begin{eqnarray}
 &  & \widetilde{G}^{\alpha\beta\rho\sigma}(p)=\frac{\kappa}{16\mu^{4}}\Biggl(16\mu^{4}\eta^{\alpha\beta}\eta^{\rho\sigma}-8\mu^{4}(\eta^{\alpha\sigma}\eta^{\beta\rho}+\eta^{\alpha\rho}\eta^{\beta\sigma})\\
 &  & ~~~~~~~~~-6\mu^{2}\kappa^{2}(\eta^{\alpha\rho}p^{\beta}p^{\sigma}+\eta^{\beta\rho}p^{\alpha}p^{\sigma}+\eta^{\alpha\sigma}p^{\beta}p^{\rho}+\eta^{\beta\sigma}p^{\alpha}p^{\rho})\nonumber \\
 &  & ~~~~~~~~+8\mu^{2}\kappa^{2}(\eta^{\alpha\beta}p^{\rho}p^{\sigma}+\eta^{\rho\sigma}p^{\alpha}p^{\beta})-6\kappa^{4}p^{\alpha}p^{\beta}p^{\rho}p^{\sigma}\nonumber \\
 &  & ~~~~~~~~-4i\mu^{3}\kappa p_{\kappa}(\eta^{\kappa\alpha\rho}\eta^{\beta\sigma}+\eta^{\kappa\beta\rho}\eta^{\alpha\sigma}+\eta^{\kappa\alpha\sigma}\eta^{\beta\rho}+\eta^{\kappa\beta\sigma}\eta^{\alpha\rho})\nonumber \\
 &  & ~~~~~~~~-3i\mu\kappa^{3}p_{\kappa}(\eta^{\kappa\alpha\rho}p^{\beta}p^{\sigma}+\eta^{\kappa\beta\rho}p^{\alpha}p^{\sigma}+\eta^{\kappa\alpha\sigma}p^{\beta}p^{\rho}+\eta^{\kappa\beta\sigma}p^{\alpha}p^{\rho})\Biggr)\frac{1}{p^{2}}\nonumber \\
 &  & ~~~~~~~~~+\frac{\kappa}{4\mu^{4}}\Biggl(-2\mu^{4}\eta^{\alpha\beta}\eta^{\rho\sigma}+2\mu^{4}(\eta^{\alpha\sigma}\eta^{\beta\rho}+\eta^{\alpha\rho}\eta^{\beta\sigma})\nonumber \\
 &  & ~~~~~~~~~+2\mu^{2}\kappa^{2}(\eta^{\alpha\rho}p^{\beta}p^{\sigma}+\eta^{\beta\rho}p^{\alpha}p^{\sigma}+\eta^{\alpha\sigma}p^{\beta}p^{\rho}+\eta^{\beta\sigma}p^{\alpha}p^{\rho})\nonumber \\
 &  & ~~~~~~~~~-2\mu^{2}\kappa^{2}(\eta^{\alpha\beta}p^{\rho}p^{\sigma}+\eta^{\rho\sigma}p^{\alpha}p^{\beta})+2\kappa^{4}p^{\alpha}p^{\beta}p^{\rho}p^{\sigma}\nonumber \\
 &  & ~~~~~~~~~+i\mu^{3}\kappa p_{\kappa}(\eta^{\kappa\alpha\rho}\eta^{\beta\sigma}+\eta^{\kappa\beta\rho}\eta^{\alpha\sigma}+\eta^{\kappa\alpha\sigma}\eta^{\beta\rho}+\eta^{\kappa\beta\sigma}\eta^{\alpha\rho})\nonumber \\
 &  & ~~~~~~~~~+i\mu\kappa^{3}p_{\kappa}(\eta^{\kappa\alpha\rho}p^{\beta}p^{\sigma}+\eta^{\kappa\beta\rho}p^{\alpha}p^{\sigma}+\eta^{\kappa\alpha\sigma}p^{\beta}p^{\rho}+\eta^{\kappa\beta\sigma}p^{\alpha}p^{\rho})\Biggr)\frac{1}{p^{2}+\mu^{2}/\kappa^{2}}\nonumber \\
 &  & ~~~~~~~~-\frac{\kappa^{3}}{16\mu^{4}}\Biggl(2\mu^{2}(\eta^{\alpha\rho}p^{\beta}p^{\sigma}+\eta^{\beta\rho}p^{\alpha}p^{\sigma}+\eta^{\alpha\sigma}p^{\beta}p^{\rho}+\eta^{\beta\sigma}p^{\alpha}p^{\rho})+2\kappa^{2}p^{\alpha}p^{\beta}p^{\rho}p^{\sigma}\nonumber \\
 &  & ~~~~~~~~+i\mu\kappa p_{\kappa}(\eta^{\kappa\alpha\rho}p^{\beta}p^{\sigma}+\eta^{\kappa\beta\rho}p^{\alpha}p^{\sigma}+\eta^{\kappa\alpha\sigma}p^{\beta}p^{\rho}+\eta^{\kappa\beta\sigma}p^{\alpha}p^{\rho})\Biggr)\frac{1}{p^{2}+4\mu^{2}/\kappa^{2}}.\nonumber 
\end{eqnarray}
Applying the inverse Fourier transformation we obtain 
\begin{eqnarray}
 &  & G^{\alpha\beta\rho\sigma}(t,r)=\frac{\kappa}{16\mu^{4}}\Biggl(16\mu^{4}\eta^{\alpha\beta}\eta^{\rho\sigma}-8\mu^{4}(\eta^{\alpha\sigma}\eta^{\beta\rho}+\eta^{\alpha\rho}\eta^{\beta\sigma})\\
 &  & +4\mu^{3}\kappa(\eta^{\kappa\alpha\rho}\eta^{\beta\sigma}+\eta^{\kappa\beta\rho}\eta^{\alpha\sigma}+\eta^{\kappa\alpha\sigma}\eta^{\beta\rho}+\eta^{\kappa\beta\sigma}\eta^{\alpha\rho})\partial_{\kappa}\Biggr)\frac{1}{2\pi}\Theta(t)\Theta(t-r)\frac{1}{\sqrt{t^{2}-r^{2}}}\nonumber \\
 &  & +\frac{\kappa}{4\mu^{4}}\Biggl(-2\mu^{4}\eta^{\alpha\beta}\eta^{\rho\sigma}+2\mu^{4}(\eta^{\alpha\sigma}\eta^{\beta\rho}+\eta^{\alpha\rho}\eta^{\beta\sigma})\nonumber \\
 &  & -\mu^{3}\kappa(\eta^{\kappa\alpha\rho}\eta^{\beta\sigma}+\eta^{\kappa\beta\rho}\eta^{\alpha\sigma}+\eta^{\kappa\alpha\sigma}\eta^{\beta\rho}+\eta^{\kappa\beta\sigma}\eta^{\alpha\rho})\partial_{\kappa}\Biggr)\frac{1}{2\pi}\Theta(t)\Theta(t-r)\frac{\cos\left(\frac{\mu}{\kappa}\sqrt{t^{2}-r^{2}}\right)}{\sqrt{t^{2}-r^{2}}},\nonumber 
\end{eqnarray}
where, in the last expression, we dropped the terms which vanish when the propagator is
sandwiched between two conserved sources. There is a single massive spin-2 mode with only one degree of freedom since it is a parity non-invariant theory.  Once again there is a tail
inside the light cone and the Huygens' principle is violated. 

\section{NEW MASSIVE GRAVITY IN de SITTER SPACETIME}

As mentioned in the introduction, anomalous dispersion can disappear
in the $2+1$ dimensional gravity in a curved background. To understand
this, let us consider a generic quadratic gravity in a de Sitter background.
This theory was studied in detail in \cite{Tekin3}. Here, we
shall only quote the pertaining details for our discussion. Generic
three dimensional quadratic action is 
\begin{equation}
I=\int d^{3}x\,\sqrt{-g}\Bigl(\frac{1}{\kappa}\left(R-2\Lambda_{0}\right)+\alpha R^{2}+\beta R_{\mu\nu}^{2}\Bigr).
\end{equation}
Consider the linearization of this theory in a de Sitter background given
by the following metric 
\begin{equation}
ds^{2}=\frac{\ell^{2}}{t^{2}}\left(-dt^{2}+dx^{2}+dy^{2}\right),
\end{equation}
where the effective cosmological constant is 
\begin{equation}
\frac{1}{\ell^{2}}=\frac{1}{4\kappa\left(3\alpha+\beta\right)}\Bigl(1\pm\sqrt{1-8\kappa\Lambda_{0}\left(3\alpha+\beta\right)}\Bigr).
\end{equation}
Defining the perturbations as 
\begin{equation}
g_{\mu\nu}=\frac{\ell^{2}}{t^{2}}\eta_{\mu\nu}+h_{\mu\nu},
\end{equation}
a rather long discussion given in \cite{Tekin3, Tekin_Rapid} one shows that for
generic $\alpha$, $\beta$, $\kappa$ there are three propagating
degrees of freedom. Two of these constitute the massive spin-$2$
field with the mass 
\begin{equation}
m_{g}^{2}=-\frac{1}{\kappa\beta}-\frac{12\alpha}{\ell^{2}\beta}-\frac{4}{\ell^{2}},\label{eq:dS_m_g}
\end{equation}
and the third degree of freedom is a spin-$0$ mode with the mass
\begin{equation}
m_{s}^{2}=\frac{1}{\kappa\left(8\alpha+3\beta\right)}-\frac{4}{\ell^{2}}\left(\frac{3\alpha+\beta}{8\alpha+3\beta}\right).\label{ms^2}
\end{equation}
Let $\varphi$ denote the spin-$0$ field, which arises as a gauge-invariant
object then its action is given as 
\begin{equation}
I_{\varphi}=\frac{\left(8\alpha+3\beta\right)}{8}\int d^{3}x\,\left[\frac{t^{3}}{\ell^{3}}\dot{\varphi}^{2}-\frac{1}{\left(8\alpha+3\beta\right)}\frac{t}{\ell}\left(\frac{1}{\kappa}-\frac{12\alpha}{\ell^{2}}-\frac{4\beta}{\ell^{2}}\right)\varphi^{2}\right].\label{phiaction}
\end{equation}
On the other hand, the two modes of the spin-$2$ field comes with
the following action 
\begin{equation}
I_{\sigma}=\frac{1}{2}\int d^{3}x\,\left[\beta\frac{t^{3}}{\ell^{3}}\left(\dot{\sigma}^{2}+\sigma\nabla^{2}\sigma\right)+\left(\frac{1}{\kappa}+\frac{12\alpha}{\ell^{2}}+\frac{4\beta}{\ell^{2}}\right)\frac{t}{\ell}\sigma^{2}\right].\label{eq:sigma_action}
\end{equation}
To understand these modes, let us recall that a free scalar field in this background with mass $m$ has the action 
\[
I=-\frac{1}{2}\int d^{3}x\,\sqrt{-g}\left(\partial_{\mu}\Phi\partial^{\mu}\Phi+m^{2}\Phi^{2}\right)=-\frac{1}{2}\int d^{3}x\,\left\{ \frac{\ell}{t}\left[-\dot{\Phi}^{2}+\left(\partial_{i}\Phi\right)^{2}\right]+\frac{\ell^{3}}{t^{3}}m^{2}\Phi^{2}\right\} .
\]
Comparing this with (\ref{phiaction}) and (\ref{eq:sigma_action}),
after scaling $\sigma$ as $\sigma\rightarrow\frac{\ell^{2}}{t^{2}}\sigma$
and similarly $\varphi\rightarrow\frac{\ell^{2}}{t^{2}}\varphi$ one
can read (\ref{ms^2}) from (\ref{phiaction}) and (\ref{eq:dS_m_g})
from (\ref{eq:sigma_action}). In the NMG limit, that is $8\alpha+3\beta=0$
the $\varphi$ field is infinitely massive and drops out of the
spectrum. One is left with a massive spin-$2$ field with the mass
\begin{equation}
m_{g}^{2}=-\frac{1}{\kappa\beta}+\frac{1}{2\ell^{2}}.\label{eq:dS_m_g-1}
\end{equation}
As shown in \cite{Tekin_meta}, a massive scalar field in
de Sitter spacetime with the tuned mass $m=1/\ell$ shows dispersionless propagation in
$2+1$ dimensions. One can easily see this from the following construction:
with the coordinate change $t=\ell e^{-\tau/\ell}$ and $a(\tau)=e^{\tau/\ell}$,
de Sitter metric becomes 
\[
ds^{2}=-d\tau^{2}+a(\tau)^{2}\Bigl(dx^{2}+dy^{2}\Bigr).
\]
In these coordinates, the Fourier modes of the massive graviton has
the dispersion relation 
\begin{equation}
w_{\vec{k}}^{2}=-\frac{1}{\ell^{2}}+ m_{g}^{2}+\frac{k^{2}}{a^{2}}.\label{dispersion}
\end{equation}
For $m_{g}^{2}=1/\ell^{2}$, the group velocity is independent of 
$\vec{k}$. Hence anomalous dispersion disappears. This
corresponds to the case $\kappa\beta=-2\ell^{2}$. This is possible for
$\Lambda_{0}=-27/\ell^{2}$. Note that, the bare cosmological constant
is negative, but the effective cosmological constant is positive. 

One can easily generalize the discussion of the previous section to generic $D$ dimensions. Consider a massive scalar field living in the background spacetime with the metric
\begin{equation}
ds^{2}=-d\tau^{2}+a(\tau)^{2} \sum_{i=1}^{D-1}dx^i d x^i, \hskip 1 cm a(\tau) = e^{H \tau} \hskip 1 cm H:= \sqrt{\frac{ 2 \Lambda}{(D-1)(D-2)}}.
\end{equation}
Then the wave equation $(\Box - m^2)\Phi=0$, is solved by the Fourier modes 
\begin{equation}
\Phi (\tau, x^i) := \frac{f_{\vec{k}}(\tau)}{a(\tau)} e^{i \vec{k}\cdot \vec{x}},
\end{equation}
as long as the following equation is satisfied
\begin{equation}
\ddot f_{\vec{k}}(\tau) + \omega_{k}^2 f_{\vec{k}}(\tau)=0, \hskip 1 cm  \omega_{k}^2 := m^2 - \frac{\Lambda (D-1)}{ 2 (D-2)} +\frac{k^2}{a(\tau)^2}. 
\label{harmonic}
\end{equation}
Note that for the tuning $m^2 =  \frac{\Lambda (D-1)}{ 2 (D-2)}$, the group velocity $ v^i_g = \frac{\partial \omega_k}{\partial k^i} $ is independent of $k$ and hence there is no dispersion. For this case, the solution to (\ref{harmonic}) is given in terms of the Bessel function of the first and second kinds as 
\begin{equation}
f_{\vec{k}}(\tau)  = c_1 J_0\left(\frac{k}{a(\tau)H}\right)+c_2 Y_0\left(\frac{k}{a(\tau)H}\right),
\end{equation}
and a generic wave pulse can be constructed from the superposition of these modes. Note that in \cite{Yag}\footnote{ We thank a conscientious referee for bringing this reference to our attention.} the same result was reached but to make the proper comparison $H=1 \rightarrow \Lambda = \frac{ (D-1)(D-2)}{2} $ choice should be made in our expressions. 

\section{Conclusions}
We studied the propagation of gravity waves in some detail in flat and de Sitter spacetimes for massless and massive gravity, quadratic gravity theories. It is quite well-known that in flat backgrounds, with odd number of spatial dimensions (such as our universe), there is no anomalous dispersion in a vacuum, while for all even spatial dimensions there is anomalous dispersion. So introducing one space-like dimension changes the propagation dramatically. What has been a rather unexpected surprise was to see that adding one time-like dimensions also removes anomalous dispersion  which was demonstrated in \cite{Bender} for a particular setting whose details have been given in \cite{Tekin_meta}. In this work we have studied the extensions of these considerations to massive gravity theories, in particular showed that for a particular tuning of the mass in terms of the cosmological constant, both scalar waves in $D$ dimensional de Sitter spacetime and new massive gravity in 2+1 dimensions allow dispersion-free propagation and hence the Huygens' principle survives.

\section*{Acknowledgments}

The works of E.A. and E.K. are partially supported by the TUBITAK
Grant No. 120F253. The work of E.K. is partially supported by the
TUBITAK Grant No. 119F241.

\appendix

\section{ Massless Integral}

Here, for completeness, we give some of the details of the integrals that we used in the body of the text. We claim no originality in these two appendices as these results can be found in various forms in the literature \cite{CH,Dick,Hassani, Soodak, Duffy}. We used \cite{Gradshteyn} for various integrals and relations.

Consider the mass free integral 
\begin{equation}
I_{1}:=\int\frac{d^{d+1}p}{(2\pi)^{d+1}}\frac{e^{-ip\cdot (x-x')}}{p^{2}},
\end{equation}
which reads 
\begin{equation}
I_{1}=\int\frac{d^{d}p}{(2\pi)^{d}}e^{-i\vec{p}\cdot\vec{r}}\int_{-\infty}^{\infty}\frac{dp^{0}}{2\pi}\frac{e^{ip^{0}t}}{\vec{p}^{2}-(p^{0})^{2}},
\end{equation}
where we have defined ${\vec{r}:={\vec{x}-{\vec{x}'}}}$ and $t:=x^{0}-x'^{0}$. To obtain the retarded Green's function, in carrying out the $p^0$ integral, both poles are displaced in such a way that they are located in the upper-half plane and contribute to the integral. 
Hence the $p^{0}$ integral yields 
\begin{equation}
\int_{-\infty}^{\infty}\frac{dp^{0}}{2\pi}\frac{e^{ip^{0}t}}{\vec{p}^{2}-(p^{0})^{2}}=\frac{\sin(pt)}{p}\varTheta(t),
\end{equation}
where $\varTheta(t)$ denotes the Heaviside step function and $p:=|\vec{p}|$. Then
\begin{equation}
I_{1}=\varTheta(t)\int\frac{d^{d}p}{(2\pi)^{d}}e^{-i\vec{p}.\vec{r}}\frac{\sin(pt)}{p}.
\end{equation}
Assuming $\vec{p}\cdot\vec{r}=\bigl|\vec{p}\bigr|\bigl|\vec{r}\bigr|\cos\theta_{1}$
and using 
\begin{equation}
d^{d}p=p^{d-1}(\sin\theta_{1})^{d-2}(\sin\theta_{2})^{d-3}...\sin\theta_{d-2}d\theta_{1}d\theta_{2}....d\theta_{d-1}dp
\end{equation}
we obtain 
\begin{equation}
I_{1}=\varTheta(t)\intop_{0}^{\infty}\frac{dp}{(2\pi)^{d}}\sin(pt)p^{d-2}\intop_{0}^{\pi}d\theta_{1}e^{-i\bigl|\vec{p}\bigr|\bigl|\vec{r}\bigr|\cos\theta_{1}}(\sin\theta_{1})^{d-2}\int d\theta_{2}....d\theta_{d-1}(\sin\theta_{2})^{d-3}...\sin\theta_{d-2},
\end{equation}
where
\begin{equation}
\int d\theta_{2}....d\theta_{d-1}(\sin\theta_{2})^{d-3}...\sin\theta_{d-2}=\frac{2\pi^{(d-1)/2}}{ \Gamma\left(\frac{d-1}{2}\right)}
\end{equation}
is the solid angle in $d-1$ dimensions for both the even and odd dimensional cases. To evaluate the $\theta_{1}$ integral we use the following formula
\begin{equation}
\intop_{0}^{\pi}d\theta e^{ikr\cos\theta}(\sin\theta)^{m-2}=\sqrt{\pi}\left(\frac{2}{kr}\right)^{(m-2)/2}\Gamma\left(\frac{m-1}{2}\right)J_{m/2-1}(kr).
\end{equation}
Then we get 
\begin{equation}
I_{1}=\varTheta(t)(2\pi)^{-d/2}\intop_{0}^{\infty}dp\thinspace\sin(pt)p^{d-2}\frac{J_{d/2-1}(-pr)}{(-pr)^{d/2-1}}.\label{massfreeintegral}
\end{equation}
To complete the calculation, we need to compute the term $J_{d/2-1}(-pr)/(-pr)^{d/2-1}$.
We use the identity \cite{Dick}
\begin{equation}
\frac{J_{v+n}(x)}{x^{v+n}}=\left(-\frac{1}{x}\frac{d}{dx}\right)^{n}\frac{J_{v}(x)}{x^{v}}.\label{identityDick}
\end{equation}
Now, let us consider the even and odd dimensional cases separately.

\subsubsection{\textbf{Odd $d$ case }}

We have $x=-pr$ in (\ref{identityDick}). Let $v=-1/2$, then one
has $n=(d-1)/2$ and we use 
\begin{equation}
J_{-1/2}(z)=\sqrt{\frac{2}{\pi z}}\cos z
\end{equation}
to arrive at 
\begin{equation}
\frac{J_{d/2-1}(-pr)}{(-pr)^{d/2-1}}=2^{d/2}\pi^{(d-2)/2}p^{1-d}\left(-\frac{1}{2\pi r}\frac{d}{dr}\right)^{(d-1)/2}\cos(pr).\label{identity1}
\end{equation}
Then $I_{1}$ integral reduces to 
\begin{equation}
I_{1}=\frac{1}{\pi}\varTheta(t)\left(-\frac{1}{2\pi r}\frac{d}{dr}\right)^{(d-1)/2}\intop_{0}^{\infty}dp\thinspace\frac{\sin(pt)\cos(pr)}{p}.
\end{equation}
In order to take the $p$ integral one needs \cite{Gradshteyn}
\begin{equation}
\intop_{0}^{\infty}dx\thinspace\frac{\sin(ax)\cos(bx)}{x}=\begin{cases}
\pi/2,~~~~~~~~~~~~~~~~~~~~~~~a>b\geq0\\
\pi/4,~~~~~~~~~~~~~~~~~~~~~~~~a\text{=}b>0\\
0,~~~~~~~~~~~~~~~~~~~~~~~~~~~b>a\geq0.
\end{cases}
\end{equation}
Using these one ends up with
\begin{equation}
I_{1}=\frac{1}{4\pi}\varTheta(t)\left(-\frac{1}{2\pi r}\frac{d}{dr}\right)^{(d-3)/2}\frac{\delta(t-r)}{r}.
\end{equation}

\subsubsection{\textbf{Even $d$ case }}

We have $x=-pr$ in (\ref{identityDick}). Let $v=0$, then one has
$n=(d-2)/2$ and (\ref{identityDick}) yields 
\begin{equation}
\frac{J_{d/2-1}(-pr)}{(-pr)^{d/2-1}}=p^{(2-d)/2}\left(\frac{1}{r}\frac{d}{d(-pr)}\right)^{(d-2)/2}J_{0}(-pr),
\end{equation}
where 
\begin{equation}
J_{0}(z)=\sum_{k=0}^{\infty}(-1)^{k}\frac{z^{2k}}{2^{2k}(k!)^{2}}.
\end{equation}
One has $J_{0}(-z)=J_{0}(z)$, then we obtain
\begin{equation}
\frac{J_{d/2-1}(-pr)}{(-pr)^{d/2-1}}=(2\pi)^{(d-2)/2}p^{2-d}\left(-\frac{1}{2\pi r}\frac{d}{dr}\right)^{(d-2)/2}J_{0}(-pr).\label{identity2}
\end{equation}
Substituting this in (\ref{massfreeintegral}) we get 
\begin{equation}
I_{1}=\frac{1}{2\pi}\varTheta(t)\left(-\frac{1}{2\pi r}\frac{d}{dr}\right)^{(d-2)/2}\intop_{0}^{\infty}dp\thinspace\sin(pt)J_{0}(-pr),
\end{equation}
and we end up with 
\begin{equation}
I_{1}=\frac{1}{2\pi}\varTheta(t)\left(-\frac{1}{2\pi r}\frac{d}{dr}\right)^{(d-2)/2}\frac{\Theta(t-r)}{\sqrt{t^{2}-r^{2}}}.
\end{equation}
We can summarize the results as follows \cite{CH, Hassani, Soodak}
\begin{equation}
I_{1}=\int\frac{d^{d+1}p}{(2\pi)^{d+1}}\frac{e^{-ip(x-x')}}{p^{2}}=\begin{cases}
\frac{1}{4\pi}\varTheta(t)\left(-\frac{1}{2\pi r}\frac{d}{dr}\right)^{(d-3)/2}\frac{\delta(t-r)}{r} :\,\text{for}\,\,\text{odd}\,\,d,\\
\\
\frac{1}{2\pi}\varTheta(t)\left(-\frac{1}{2\pi r}\frac{d}{dr}\right)^{(d-2)/2}\frac{\Theta(t-r)}{\sqrt{t^{2}-r^{2}}} :\,\text{for}\,\,\text{even}\,\,d.
\end{cases}
\end{equation}

\section{Massive integral}

Let us consider the following integral
\begin{equation}
I_{2}:=\int\frac{d^{d+1}p}{(2\pi)^{d+1}}\frac{e^{-ip\cdot(x-x')}}{p^{2}+m^{2}}.
\end{equation} 
Similar steps in the previous section yields 
\begin{equation}
I_{2}=\varTheta(t)(2\pi)^{-d/2}\intop_{0}^{\infty}dpp^{d-1}\frac{\sin(t\sqrt{\vec{p}^{2}+m^{2}})}{\sqrt{\vec{p}^{2}+m^{2}}}\frac{J_{d/2-1}(-pr)}{(-pr)^{d/2-1}}\label{massincludingintegral}.
\end{equation}
Now we need to consider the odd and even $d$ separately.

\subsubsection{\textbf{Odd $d$ case }}

Using the identity (\ref{identity1}) the $I_{2}$ integral reduces
to 
\begin{equation}
I_{2}=\frac{1}{\pi}\varTheta(t)\left(-\frac{1}{2\pi r}\frac{d}{dr}\right)^{(d-1)/2}\intop_{0}^{\infty}dp\frac{\sin(t\sqrt{\vec{p}^{2}+m^{2}})}{\sqrt{\vec{p}^{2}+m^{2}}}\cos(pr).
\end{equation}
In order to take the $p$ integral we use 
\begin{equation}
\intop_{0}^{\infty}dx\thinspace\frac{\sin\bigl(p\sqrt{x^{2}+a^{2}}\bigr)\cos(bx)}{x^{2}+a^{2}}=\begin{cases}
\pi J_{0}\bigl(a\sqrt{p^{2}-b^{2}}\bigr)/2,~~~~~~~~~~~~~0<b<p,~~~a>0\\
0,~~~~~~~~~~~~~~~~~~~~~~~~~~~~~~~~~~~~b>p>0,~~\ a>0
\end{cases}
\end{equation}
and we arrive at
\begin{equation}
I_{2}=\frac{1}{2}\varTheta(t)\left(-\frac{1}{2\pi r}\frac{d}{dr}\right)^{(d-1)/2}\Theta(t-r)J_{0}\bigl(m\sqrt{t^{2}-r^{2}}\bigr).
\end{equation}

\subsubsection{\textbf{Even $d$ case }}

Using (\ref{identity2}) we get
\begin{equation}
I_{2}=\frac{1}{2\pi}\varTheta(t)\left(-\frac{1}{2\pi r}\frac{d}{dr}\right)^{(d-2)/2}\intop_{0}^{\infty}dp\frac{p\sin(t\sqrt{\vec{p}^{2}+m^{2}})}{\sqrt{\vec{p}^{2}+m^{2}}}J_{0}(pr),
\end{equation}
where 

\begin{equation}
\int_{0}^{\infty}dpJ_{0}(pr)\frac{p\sin(t\sqrt{\vec{p}^{2}+m^{2}})}{\sqrt{\vec{p}^{2}+m^{2}}}=\frac{\cos(m\sqrt{t^{2}-r^{2}})}{\sqrt{t^{2}-r^{2}}}\Theta(t-r).\label{integralSergej-1-1}
\end{equation}
Then one obtains
\begin{equation}
I_{2}=\frac{1}{2\pi}\varTheta(t)\left(-\frac{1}{2\pi r}\frac{d}{dr}\right)^{(d-2)/2}\frac{\cos(m\sqrt{t^{2}-r^{2}})}{\sqrt{t^{2}-r^{2}}}\Theta(t-r).
\end{equation}
We can summarize the results as follows \cite{Duffy}
\begin{equation}
I_{2}=\int\frac{d^{d+1}p}{(2\pi)^{d+1}}\frac{e^{-ip(x-x')}}{p^{2}+m^{2}}=\begin{cases}
\frac{1}{2}\varTheta(t)\left(-\frac{1}{2\pi r}\frac{d}{dr}\right)^{(d-1)/2}\Theta(t-r)J_{0}\bigl(m\sqrt{t^{2}-r^{2}}\bigr):\,\text{for}\,\,\text{odd}\,\,d,\\
\\
\frac{1}{2\pi}\varTheta(t)\left(-\frac{1}{2\pi r}\frac{d}{dr}\right)^{(d-2)/2}\frac{\cos(m\sqrt{t^{2}-r^{2}})}{\sqrt{t^{2}-r^{2}}}\Theta(t-r):\,\text{for}\,\,\text{even}\,\,d.
\end{cases}
\end{equation}


\begin{thebibliography}{1}

\bibitem{Tekin_meta}
B.~Tekin,
Hyperbolic metamaterials and massive Klein-Gordon equation in (2+1)-dimensional de Sitter spacetime,
Phys. Rev. D \textbf{104}, no.10, 105004 (2021).

\bibitem{CH} R.~Courant and D.~Hilbert, {\it Methods of Mathematical Physics},
Vol.~2 (Wiley, New York, 1962).

\bibitem{Bender} C. M. Bender and F. J. Rodr{\'i}guez-Fortu{\~n}o, S. Sarkar and A. V. Zayats, Two-Dimensional Pulse Propagation without Anomalous Dispersion,
  Phys. Rev. Lett.~{\bf 119},
114301 (2017).

\bibitem{Bars} I. Bars and J. Terning, {\it Extra dimensions in space and
time} (Springer, New York, 2010).

\bibitem{SN} I.~I.~Smolyaninov and E.~E.~Narimanov, Metric signature transitions in optical metamaterials
Phys. Rev. Lett.~{\bf 105},
067402 (2010).

\bibitem{Sm2}
I.~I.~ Smolyaninov, Hyperbolic metamaterials,
arXiv:1510.07137 [physics.optics].

\bibitem{Tekin_Rapid} B.~Tekin, Particle Content of Quadratic
and $f(R_{\mu\nu\sigma\rho})$ Theories in $(A)dS$, Phys. Rev.
D \textbf{93}, no.10, 101502 (2016).


\bibitem{Stelle}
K.~S.~Stelle,
Renormalization of Higher Derivative Quantum Gravity,
Phys. Rev. D \textbf{16}, 953-969 (1977).

\bibitem{nmg1}
E.~A.~Bergshoeff, O.~Hohm and P.~K.~Townsend,
Massive Gravity in Three Dimensions,
Phys. Rev. Lett. \textbf{102}, 201301 (2009).

\bibitem{nmg2}
E.~A.~Bergshoeff, O.~Hohm and P.~K.~Townsend,
More on Massive 3D Gravity,
Phys. Rev. D \textbf{79}, 124042 (2009).

\bibitem{Gullu}
I.~Gullu and B.~Tekin,
Massive Higher Derivative Gravity in D-dimensional Anti-de Sitter Spacetimes,
Phys. Rev. D \textbf{80}, 064033 (2009).

\bibitem{djt} S.~Deser, R.~Jackiw and S.~Templeton, Annals Phys.\ \textbf{140},
372 (1982); Phys.\ Rev.\ Lett.\ \textbf{48}, 975 (1982).

\bibitem{Tekin3}
I.~Gullu, T.~C.~Sisman and B.~Tekin,
Canonical structure of higher derivative gravity in 3D,''
Phys. Rev. D \textbf{81}, 104017 (2010).

\bibitem{Yag} K.~Yagdjian,
``Huygens' Principle for the Klein-Gordon equation in the de Sitter spacetime,''
J. Math. Phys. \textbf{54}, 091503 (2013).

\bibitem{Dick} R. Dick, {\it Advanced quantum mechanics: materials and photons},  Springer (2018).

\bibitem{Hassani}
S.Hassani, {\it Mathematical Physics: A Modern Introduction to Its Foundations}, Springer (1960).


\bibitem{Soodak}
H.~Soodak  and S. ~T. ~Martin ,
Wakes and waves in $N$ dimensions, American Journal of Physics \textbf{61}, 395-401 (1993).

\bibitem{Duffy}
D. ~G. ~Duffy,  {\it Green's functions with applications}, 2nd ed. (CRC Press, 2015).
 
\bibitem{Gradshteyn}
I. S. Gradshteyn and I. M. Ryzhik,  {\it Table of Integrals, Series, and Products }, 7th ed. (Elsevier/Academic Press, Amster-
dam, 2007).

\end{thebibliography}
\end{document}